# Observing the sky at extremely high energies with the Cherenkov Telescope Array: Status of the GCT project

**Sol Helene[1]**
*LUTH, Observatoire de Paris, CNRS, PSL Research University,*
*Place J. Janssen, 92195, Meudon, France*
*E-mail:* `helene.sol@obspm.fr`

**Greenshaw Tim**
*Oliver Lodge Laboratory, Liverpool University, Liverpool L69 7ZE, England*
*E-mail:* `green@liverpool.ac.uk`

**Le Blanc Oriane**
*GEPI, Observatoire de Paris, CNRS, PSL Research University*
*Place J. Janssen, 92195, Meudon, France*
*E-mail:* `oriane.le-blanc@obspm.fr`

**White Richard, for the CTA GCT project**
*MPI für Kernphysik, Postf. 103980, D 69029 Heidelberg, Germany*
*E-mail:* `richard.white@mpi-hd.mpg.de`
*CTA:* `www.cta-observatory.org`

The Cherenkov Telescope Array is the main global project of ground-based gamma-ray astronomy for the coming decades. Performance will be significantly improved relative to present instruments, allowing a new insight into the high-energy Universe [1]. The nominal CTA southern array will include a sub-array of seventy 4 m telescopes spread over a few square kilometers to study the sky at extremely high energies, with the opening of a new window in the multi-TeV energy range. The Gamma-ray Cherenkov Telescope (GCT) is one of the proposed telescope designs for that sub-array. The GCT prototype recorded its first Cherenkov light on sky in 2015. After an assessment phase in 2016, new observations have been performed successfully in 2017. The GCT collaboration plans to install its first telescopes and cameras on the CTA site in Chile in 2018-2019 and to contribute a number of telescopes to the subsequent CTA production phase.

[1]Speaker

## 1.    Introduction

The international Cherenkov Telescope Array (CTA) project will provide the scientific community with the next-generation ground-based gamma-ray instrument, allowing exploration of cosmic radiation in the very high energy (VHE) range from a few tens of GeV to 300 TeV [1][2]. Southern and northern telescope arrays will be built in Chile near Cerro Paranal, and in the Canary Island of La Palma, Spain, providing access to the whole sky. The baseline layout of the southern array in Chile is composed of imaging atmospheric Cherenkov telescopes (IACTs) of three different sizes: 4 large-size telescopes (LSTs - mirror diameter 23 m), 25 medium-size telescopes (MSTs – 12 m) and 70 small-size telescopes (SSTs – 4 m). CTA will provide a significant improvement in performance with respect to present instruments, especially with an increase by a factor of ten of the sensitivity and a wider spectral range, allowing new insight into the high-energy Universe to be gained. The production phase should span from 2018 to 2023, with routine user operation expected to start around 2022 and for about 30 years.

The opening of a new window in the multi-TeV energy range is the task of the SSTs, making possible the search for PeVatrons in our Galaxy and in the Large Magellanic Cloud (LMC), the study of hadronic and leptonic emissions from Active Galactic Nuclei (AGN) at extreme energies, and the exploration of fundamental physics topics such as Lorentz Invariance Violation (LIV).

The GCT (Gamma-ray Cherenkov Telescope) is one of the telescopes proposed for the SST section of CTA. Based on a dual-mirror Schwarzschild-Couder design [2], it will be equipped with a compact high-energy camera (CHEC). The GCT prototype installed in Meudon, France, and equipped with the camera prototype was the first one within CTA to record its first Cherenkov light on sky in 2015 [3]. An assessment phase followed to validate the instrument relatively to the CTA specifications, and a new observation campaign was scheduled in spring 2017. The GCT collaboration, including teams from Australia, France, Germany, Japan, the Netherlands, and the United-Kingdom, plans to install its first telescopes and cameras on site in Chile in 2018-2019, during the CTA pre-production phase, and to contribute a number of telescopes to the subsequent CTA production phase. This report describes the current status of the GCT, results from tests of the telescope and camera and from the recent observation campaign, and the scientific perspectives for the coming years.

## 2.    Perspectives on multi-TeV sciences

Our cosmos appears to harbor many sources emitting in the highest part of the electromagnetic spectrum, above tens of TeV, as proved by large field of view instruments like MILAGRO and HAWC. Extrapolation of current data from IACTs like HESS, MAGIC and VERITAS also provides a variety of phenomena to be studied at such energies [1]. However this extreme electromagnetic cosmic window is still poorly known due to the current instrumental limitation in sensivity and angular resolution. CTA and its large sub-array of 4 m SSTs will allow for the first time a deep exploration of this domain from 5 to 300 TeV, combining the guarantee of important astrophysical results with a discovery potential in cosmology and fundamental physics.

---

[2] www.cta-observatory.org

The search for cosmic accelerators of particles at PeV energies, the so-called "PeVatrons", in our Galaxy and in the LMC, will be a key area for the SST sub-array, with the long-standing issue of the origin of the galactic cosmic rays and their influence on the interstellar environment. The discovery of a power-law spectrum without any cutoff or break, up to tens of TeV, from the diffuse emission within the central 10 parsecs of the Milky Way provided a first evidence of a cosmic hadronic PeVatron in the Galactic Center [4]. CTA will allow to further investigate the origin of such potential PeVatron and its possible causal link with the galactic central black hole Sagittarius A* (see Fig.1). Powerful VHE sources have also been detected in the LMC, and especially 30DorC, the first unambiguous detection of a superbubble in the TeV range. The 30DorC complex exhibits extreme conditions and could represent another type of PeVatron to study with CTA [6].

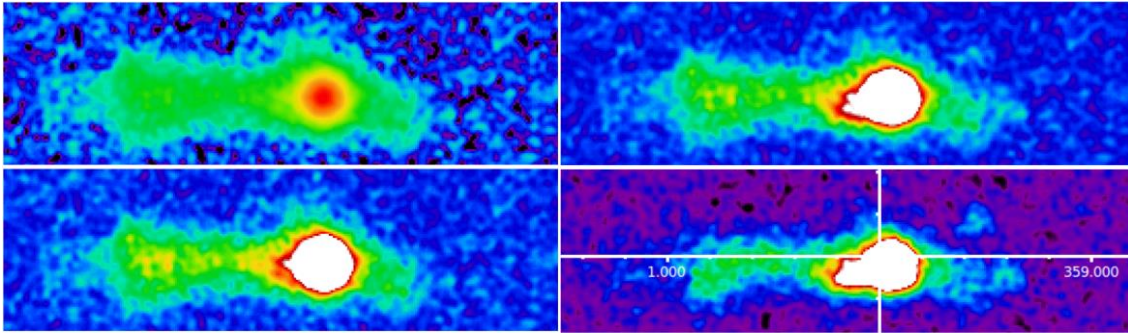

Figure 1: A simulated view of the Galactic Centre region for various astrophysical scenarios (excess events above 800 GeV after cosmic background substraction and Gaussian smoothing). On the left: 2point-like sources for SgrA* and SNR G0.9+0.1, plus diffuse emission (top image in log scale, lower image clipped). On the right: same model with a catalogue of pulsar wind nebulae (PWN) in addition (top image), and alternative model with extended circum-nuclear ring gas interacting with isotropic cosmic rays; central source and G0.9+0.1 excluded (lower image). From [5].

The multi-TeV range will also explore the most powerful acceleration mechanisms in nearby AGN and in AGN flares, and search for signatures of hadronic versus leptonic processes in such extragalactic sources [7,8]. Studying the propagation along the line of sight of extremely high energy photons from AGN will offer the opportunity to analyze the diffuse extragalactic background light (EBL) and the intergalactic magnetic field, and to probe the fine structure of spacetime, looking for potential clues of LIV [9]. Several versions of quantum gravity theories imply that LIV can significantly reduce the EBL opacity to gamma-rays above 10 TeV. The detection of such anomalies in the cosmic opacity could be reachable by CTA with the search for LIV upturn in the multi-TeV spectra of bright blazars [10]. Extending the spectral range up to extreme energies should also facilitate the studies of arrival time delays of VHE photons with their energy, which is another possible signature of LIV phenomena [11].

## 3.  The GCT project and its prototype

The GCT is an alt-azimuthal dual-mirror telescope based on a Schwarzschild-Couder (SC) optical design never built in astronomy before the advent of CTA. Such SC design offers many advantages for ground-based gamma-ray astronomy, with large field of view and reduced focal length allowing compact and lightweight telescope and cameras equipped with Multianode Photo-multiplier (MAPM) and Silicon Photo-multiplier (SiPM) detectors. To test the final performance of the SC design and validate studies and processes, a complete prototype of the GCT telescope with fast electronics cameras, one MAPM and one SiPM camera, has been implemented and its performances simulated by Monte Carlo technics. The Cherenkov light recorded by the GCT prototype on November 26th, 2015, was the first ever recorded by a Schwarzschild-Couder telescope, never achieved before [3]. Since then further developments, implementations and tests were performed.

The mechanical telescope structure of the GCT prototype has been designed in order to facilitate production, transport, assembly and maintenance. It is composed of a telescope tower, an alt-azimuth mount, an optical support structure and a counterweight (Fig.2). The 4 m primary mirror is segmented in six aspherical petals, the 2 m secondary mirror is monolithic. The Telescope control system which controls the drives and safety systems enables local and remote operations and monitoring of the telescope. The remote control via an OPC-UA interface was implemented in early 2017 with the integration of the pointing and tracking software in the prototype embedded PC.

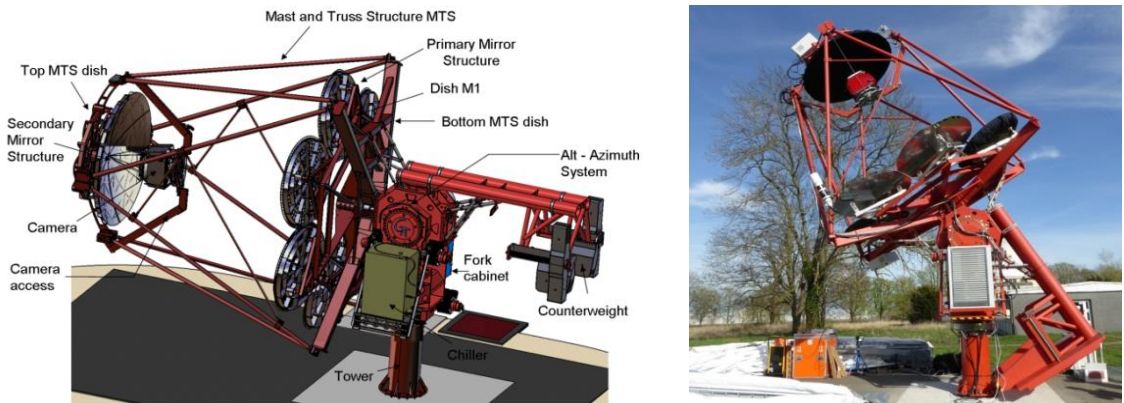

Figure 2: The GCT prototype. Left: design of the prototype showing its various sub-systems. Right: the prototype during a test in Meudon (March 2017).

The GCT prototype has undergone intensive analyses and testing in 2016 and 2017 and no technical difficulties or specific risks have been identified. The most constraining mechanical tests concern the motion of the telescope. Several quantitative measurements have been performed and some results are shown in Fig.3. The GCT mechanical design fulfils easily the current CTA requirements: the maximum possible speed allows the telescope to reach any point in the sky within 60 s and the drives power consumption during slewing remains well below the CTA power consumption maximum limit of 10 kW.

Moreover, progresses were also made in optimising the primary mirror alignment, safety and operation procedures implementation, optics characterisation, reliability, availability, maintainability and safety analysis, as well as pointing and tracking performance tests [12]. To optimise the pointing precision, an ATIK monochromatic camera was installed at the focal surface in place of the Cherenkov camera. A series of pointing data of stars provided a first model of the 2017 telescope geometry based on the TPOINT software [13]. Such model allows the pointing precision to surpass the CTA requirement, with a precision well below 210" during favourable observation conditions, before applying any predictive model for the deformations due to gravity and wind, and before any improvement due to monitoring with guide cameras or other real time monitoring devices. This should finally ensure a rms space-angle post-calibration pointing precision smaller than 7".

For more descriptive details on the tests results of the GCT structure please refer to [12].

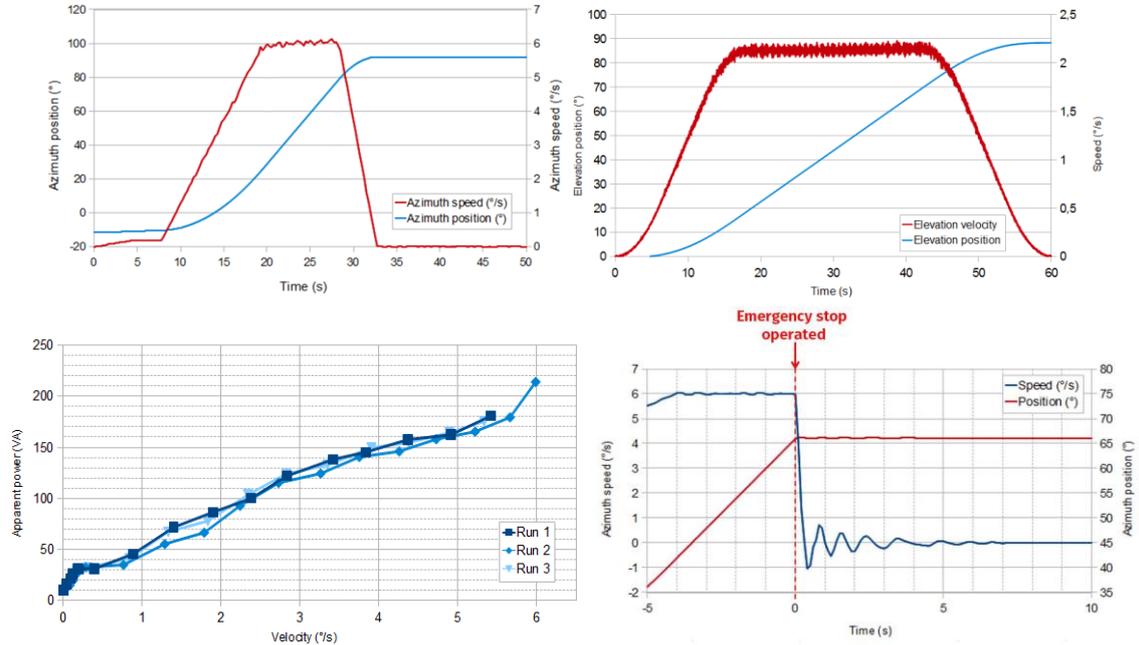

Figure 3: Mechanical and drives performances. Upper panel: azimuth and elevation positions (°) and velocities (°/s). Lower panel: apparent power (VA) versus velocity (°/s) of the azimuth axis, at 20° in elevation (left), and test of emergency stop (right) .

The CHEC camera contains 2048 pixels of about 6x6 $mm^2$, arranged on a curved focal surface to approximate the required radius of curvature requested by the SC optics. This results in a camera diameter of about 40 cm which offers a field of view of over 8 degrees with GCT. The camera electronics can sample incoming signals at 1 GSa/s and provide a flexible triggering scheme. Two prototypes have been developed, the CHEC-M equipped with MAPM [14] and the CHEC-S with SiPM tiles [15]. The CHEC-M front-end electronics (FEE) modules are based on the TARGET 5 ASICs, 16-channel devices combining digitisation and triggering functionalities [16]. A new generation of such ASICs with significantly better performances is now used in CHEC-S (see Fig.4a).

The CHEC-M and its components have undergone extensive laboratory testing. It has been also installed on the GCT prototype in Meudon for two observing campaigns. Major tests of the CHEC-M are now nearly concluded but routine operation will continue in Meudon with the help of remote camera support to better investigate the stability and reliability of the instrument (see Fig.4b). The results obtained matched performance expectations and proved the feasibility of the project. They also allowed to identify some limiting factors where further developments will be fruitful.

The CHEC-S has been designed in order to tackle such limiting factors. All its components are currently under intensive lab tests and the camera should be completed in the coming months. It will in particular benefit from the huge progress in the TARGET modules and from the rapidly evolving SiPM technology (see Fig.7).

For more descriptive details on the CHEC camera and tests results of the prototypes and components, please refer to [15].

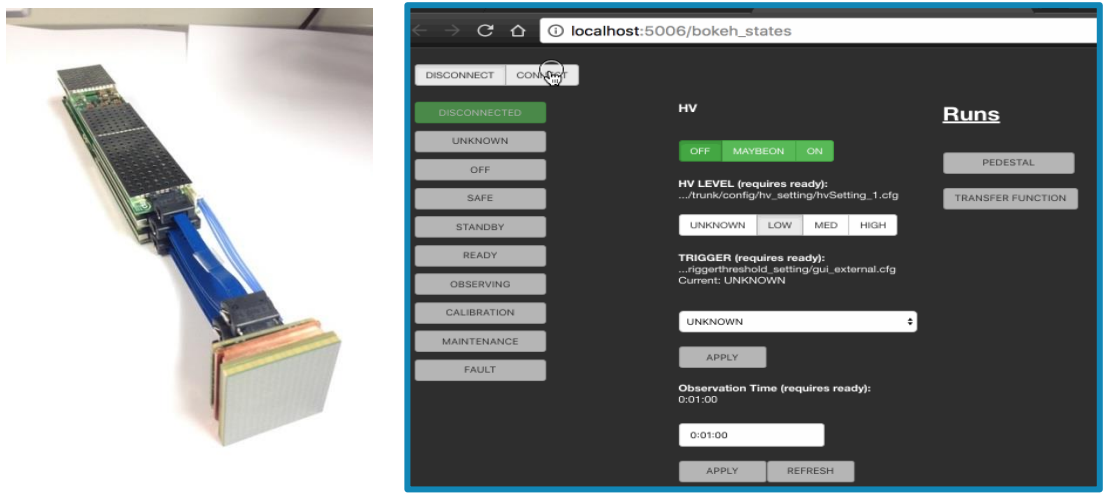

Figure 4: (a) Left panel - CHEC-S front-end electronic (FEE) module connect to each SiPM photosensor, providing full-waveform digitisation for every channel and the first-level of camera trigger, from [15]. (b) Right panel - Current camera control GUI on the GCT prototype site, with the various actions, setups and observation modes implemented.

## 4.     Preliminary results from the 2017 observing campaign

The second observation campaign of the GCT prototype took place around the new moons of March and April 2017. A series of settings adjustement, tests, and data taking were performed both for the telescope and the CHEC-M camera that were both operated remotely from the control room for the first time. Several thousands of cosmic showers were registered on the night sky under various conditions, at different azimuth and elevation without tracking, and while tracking two VHE sources, Mrk 421 and Mrk 501 (see Fig.5a, and references [12,15] for more details). The data analysis is now in progress, using when possible the standard CTA pipeline currently under development. First comparisons between gathered Cherenkov data and Monte Carlo simulations show a good agreement between the expected instrument performance and on-sky observation achievement (Fig.5b).

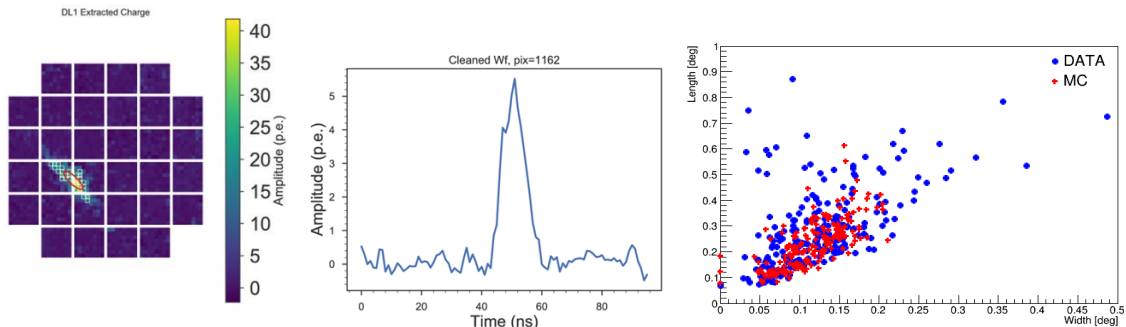

Figure 5: (a) Left panel – Camera image of one Cherenkov event and waveform of the brigtest pixel in that image, showing its time evolution over nanosecondes (middle). (b) Right panel - Comparing Cherenkov data and Monte Carlo simulations: Distribution of the cosmic showers registered during one observing run in Meudon (March 30th, 2017) in the plane "length-width" of their images (blue dots) and of the showers generated for the same setup using CORSIKA and sim_telarray simulation tools (red crosses).

## 5. Towards the construction of CTA

The GCT consortium intends to build in the coming years at least 25 telescopes equipped with Cherenkov cameras, as in-kind contributions to the CTA Observatory. Based on knowledge and expertise gained during the prototyping and assessment phases, the final GCT design of the telescope structure has been upgraded and optimised for the mass production (see Fig.6). Concerning the camera, the final design selected for the SSTs of CTA will be rather similar to the CHEC-S prototype, keeping basically the same components except for the photosensors which may benefit from significant improvements in the SiPM technology (see Fig.7). The launching of the fabrication of the GCT-01, proposed as the first GCT telescope to be built for the CTA and installed in Chile during the CTA pre-production phase, should now be imminent, to contribute to the initial partial array of CTA expected to start operating on the southern chilean site with science verification and early science in 2018-2019.

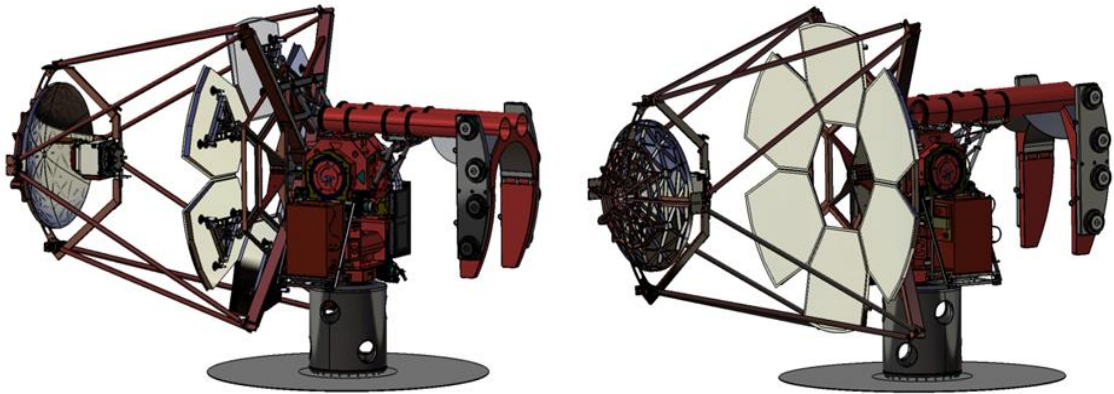

Figure 6: GCT final design proposed for the SST sub-array of CTA

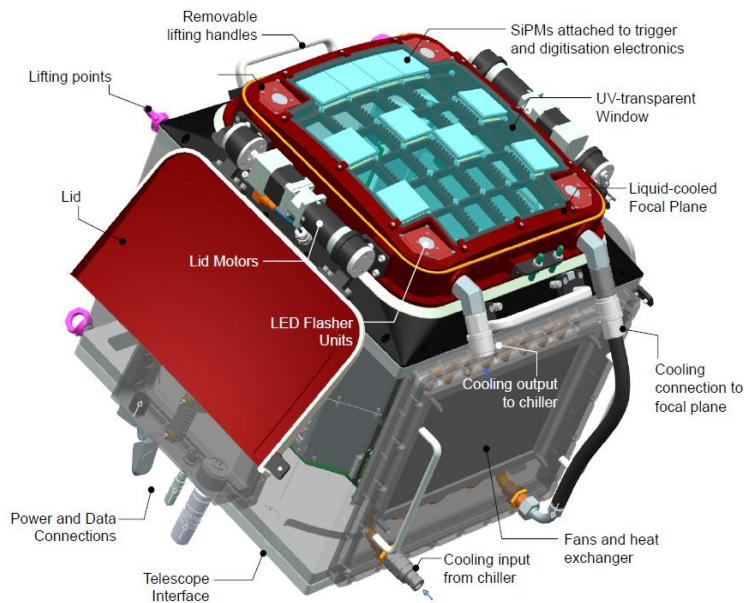

Figure 7 : General view of the CHEC-S camera proposed for the SST sub-array of CTA.

## Acknowledgments

This work was conducted in the context of the CTA Consortium. We gratefully acknowledge support from the agencies and organizations listed under Funding Agencies at this website: http://www.cta-observatory.org/consortium_acknowledgments